\documentclass[11pt]{amsart}

\usepackage[T1]{fontenc}
\usepackage{ae,aecompl}
\usepackage{amssymb,amscd}
\usepackage{url}
\usepackage{amsthm,amsfonts,amsmath,amsxtra}
\usepackage{bbm}
\usepackage{graphicx}
\usepackage{forest} 
\usepackage{epsfig,psfrag}

\numberwithin{equation}{section}
\numberwithin{table}{section}
\numberwithin{figure}{section}

\newtheoremstyle{bold}
{.5\baselineskip}{.5\baselineskip}{\itshape}{}{\bfseries}{.}{.5em}{}
\newtheoremstyle{shy}
{.5\baselineskip}{.5\baselineskip}{}{}{\bfseries}{.}{.5em}{}

\makeatletter
\def\@captionfont{\small}

\def\mychapter{%
  \if@openright\cleardoublepage\else\clearpage\fi
 \thispagestyle{empty}\global\@topnum\z@
  \@afterindenttrue \secdef\@mychapter\@schapter}

\def\@mychapter[#1]#2#3{\refstepcounter{chapter}%
  \ifnum\c@secnumdepth<\z@ \let\@secnumber\@empty
  \else \let\@secnumber\thechapter \fi
  \typeout{\chaptername\space\@secnumber}%
  \def\@toclevel{0}%
  \ifx\chaptername\appendixname \@tocwriteb\tocappendix{chapter}{#2\\ \scshape #3}%
  \else \@tocwriteb\tocchapter{chapter}{#2\\ \scshape #3}\fi
  \chaptermark{#1}%
  \addtocontents{lof}{\protect\addvspace{10\p@}}%
  \addtocontents{lot}{\protect\addvspace{10\p@}}%
  \@mymakechapterhead{#2}{#3}\@afterheading}

\def\@mymakechapterhead#1#2{\global\topskip 7.5pc\relax
  \begingroup
  \fontsize{\@xivpt}{18}\bfseries\centering
    \ifnum\c@secnumdepth>\m@ne
      \leavevmode \hskip-\leftskip
      \rlap{\vbox to\z@{\vss
          \centerline{\normalsize\mdseries
              \uppercase\@xp{\chaptername\ \thechapter}}
          \vskip 3pc}}\hskip\leftskip\fi
     #1\par \vskip 1pc
     \Large\mdseries\scshape\centering
     #2\par \endgroup
  \skip@34\p@ \advance\skip@-\normalbaselineskip
  \vskip\skip@ }

\def\section{\@startsection{section}{1}%
  \z@{.9\linespacing\@plus\linespacing}{.5\linespacing}%
  {\large\bfseries\boldmath\centering}}

\def\subsection{\@startsection{subsection}{2}%
  \z@{.7\linespacing\@plus\linespacing}{.5\linespacing}%
  {\normalfont\scshape\centering}}

\def\theindex{\@restonecoltrue\if@twocolumn\@restonecolfalse\fi
  \columnseprule\z@ \columnsep 35\p@
  \@indextitlestyle
  \thispagestyle{empty}%
  \let\item\@idxitem
  \parindent\z@  \parskip\z@\@plus.3\p@\relax
  \raggedright
  \hyphenpenalty\@M
  \footnotesize}

\renewcommand{\@bibtitlestyle}{%
  \@xp\section\@xp*\@xp{\bibname}%
}
\renewcommand{\tocchapter}[3]{%
  \indentlabel{\@ifnotempty{#2}{\ignorespaces#1 #2.\quad}}#3}

\renewcommand{\tocsection}[3]{%
  \indentlabel{\@ifnotempty{#2}{\makebox[3.2em][l]{\ignorespaces#1 #2.}}}#3}

\makeatother

\renewcommand{\tocappendix}[3]{%
  \indentlabel{#1.\quad}#3}
\renewcommand{\tocappendix}[3]{%
  \indentlabel{\makebox[5.7em][l]{\ignorespaces#1.}}#3}

\renewcommand{\bibname}{References}
\renewcommand{\ge}{\geqslant}
\renewcommand{\geq}{\geqslant}
\renewcommand{\le}{\leqslant}

\theoremstyle{bold}
\newtheorem{theorem}{Theorem}[section]
\newtheorem{proposition}[theorem]{Proposition}

\newtheorem{corollary}[theorem]{Corollary}

\theoremstyle{shy}
\newtheorem{definition}[theorem]{Definition}
\newtheorem{remark}[theorem]{Remark}

\newcommand{\EE}{\mathbb{E}}

\newcommand{\dd}{\ts\mathrm{d}\ts}
\newcommand{\ee}{\ts\mathrm{e}\ts}

\newcommand{\ts}{\hspace{0.5pt}}

\usepackage{accents}

\makeindex

\begin{document}

\title{Microbial populations under selection}

\author{Ellen Baake and Anton Wakolbinger}

\address{Faculty of Technology, Bielefeld University, 
\hspace*{\parindent}Postbox 100131, 33501 Bielefeld, Germany}
\email{ebaake@techfak.uni-bielefeld.de}

\address{Goethe University, Institute of Mathematics, 
\hspace*{\parindent} 60629 Frankfurt am Main, Germany}
\email{wakolbin@math.uni-frankfurt.de}
\maketitle

This chapter gives a synopsis of recent approaches to model and analyse the evolution of microbial populations under selection. The first part reviews two population genetic models of Lenski's long-term evolution experiment with Escherichia coli, where models aim at explaining the observed 
curve of the evolution of the mean fitness.  The second part describes a model of a host-pathogen system where the population of pathogenes experiences balancing selection, migration, and mutation, as motivated by observations of the genetic diversity of HCMV (the human cytomegalovirus) across hosts.

\section[Introduction]{Introduction}
\label{EBAW-introduction}
The genetic diversity and evolution of microbial populations is a rich and diverse object of biological research. Among the very first investigations was the  \emph{Luria-Delbr\"uck experiment} \cite{EBAW-LuDe43}, which revealed fundamental insight into the spontaneous nature of mutations,  even before the discovery of DNA as the carrier of genetic information (see \cite{EBAW-Ba08} for a historical account). \emph{Experimental evolution} takes advantage of the short generation time of bacteria and allows to observe evolution in real time; one of the most famous instances  is \index{Lenski's long-term evolution experiment} \emph{Lenski's long-term evolution experiment (LTEE)}; see \cite{EBAW-WRL13} and references therein. The 
diversity and
 evolution of \emph{pathogens} has immediate medical relevance, as exemplified by the prediction of the yearly influenza strain; see  \cite{EBAW-MoLaetal17} for a recent review. 

As also noted  in the contribution of Backofen and Pfaffelhuber \cite{EBAW-RBPP20} in this volume, population-genetic methods have so far rarely been applied to microorganisms. That chapter discusses the  diversity of the \index{CRISPR-Cas} \emph{bacterial CRISPR-Cas} (clustered regularly interspaced short palindromic repeats--CRISPR associated sequences) system, which plays an important role in the defense of  bacteria against bacteriophages. 

The present chapter reviews two recent approaches to model and analyse, in a population-genetic framework,  the evolution of microbial population under selection in combination with migration and/or mutation.  We first consider  Lenski's LTEE with \index{Escherichia coli} \emph{Escherichia coli}, that is, experimental evolution under \index{selection!directional} directional selection and mutation, and then turn to  pathogen evolution under \index{selection!balancing} balancing selection, \index{migration} migration, and mutation, as motivated by observations of the genetic diversity of \index{HCMV} \emph{HCMV} (the \index{human cytomegalovirus} \emph{human cytomegalovirus}) across hosts. For models of host-pathogen \emph{coevolution}, we refer the reader to the contribution of Stephan and Tellier \cite{EBAW-WSAT20} in this volume.

These scenarios and the corresponding models appear to be quite different  at first sight, but they share a similar spirit, for various reasons. First, due to the short generation times of microbes, it makes sense to also consider and observe the \emph{dynamics} of evolution, whereas, with higher organisms, one is often restricted to equilibrium considerations. Second, due to the large population sizes, laws of large numbers are appropriate in places, so that, under a suitable scaling, the evolution is close to a deterministic one.  Third, in both cases, a \index{weak mutation--moderate selection} \index{weak migration--moderate selection}\emph{moderate selection--weak mutation/migration} regime is used, which allows for \index{time-scale separation} time-scale separation. 

The parallels between the two parts go even further: While  the LTEE model features a process of beneficial mutations only a few of which are successful (in the sense that they lead to  fixation of the mutant),  the model for HCMV contains a process of reinfections only a few of which are effective (in the sense that they lead to a transition from a pure state to a (quasi-) equilibrium state for the \index{selection!balancing} balancing selection); we will refer to such a transition as a \emph{balancing event}. In both parts, \index{Haldane's formula} \emph{Haldane's formula} for the \index{fixation!probability} fixation probability (and its analogue in the case of \index{selection!balancing} balancing selection) plays an important role, and jump processes appear in an appropriate scaling limit, with sequential fixations in the first and sequential balancing events in the second part.

\section[Lenski's long-term evolution experiment]{Modelling Lenski's long-term evolution experiment}\label{EBAW-sec1}
\begin{figure}
  \begin{center}
 \resizebox{.4\textwidth}{!}{
%
%

\forestset{
  mytree/.style={
    for tree={
      edge+={thick},
      edge path'={
        (!u.parent anchor) -| (.child anchor)
      },
      grow=north,
      scale=0.7,
      parent anchor=children,
      child anchor=parent,
      anchor=base,
      l sep=5pt,
      s sep=10pt,
      if n children=0{align=center, base=bottom}{coordinate}
    }
  }
}

\begin{forest}
mytree
[,phantom, edge=dotted, 
 [, edge=dotted
  [, l = 10pt, edge=dotted, draw, circle, fill=black
    [, l=33pt
      [,tier=word]
      [, l=43pt
        [, l=13pt
          [,tier=word]
          [,tier=word]
        ]
        [,tier=word, draw, circle, fill=black
          [, edge=dotted, draw, circle, fill=black
            [, tier=word2]
          ]
        ]
      ]
    ]
  ]
 ]
 [, edge=dotted
  [, l = 10pt, edge=dotted, draw, circle, fill=black
    [
      [, l = 8pt
        [,l = 1pt, tier=word]
        [,l = 1pt, tier=word]
      ]
      [
        [, l = 28pt
          [,tier=word, draw, rectangle, minimum width=9pt, minimum height=9pt, fill=black
            [, edge=dotted, draw, rectangle, minimum width=9pt, minimum height=9pt, fill=black
              [, l = 15pt 
              	[, l = 1pt, tier=word2]
              	[, l = 1pt, tier=word2]
              ]
            ]
          ]
          [,tier=word]
        ]
        [, l = 18pt
          [,tier=word]
          [,tier=word]
        ]
      ]
    ]
  ]
 ]
 [, edge=dotted
  [,l = 10pt, edge=dotted, draw, circle, fill=black
    [, l=70pt
      [
        [,tier=word]
        [,tier=word]
      ]
      [,tier=word]
    ]
  ]
 ]
 [, edge=dotted
  [,l = 10pt, edge=dotted, draw, circle, fill=black
    [, l = 4pt
      [, l = 34pt
        [,tier=word, draw, circle, fill=black
          [, edge=dotted, draw, circle, fill=black
            [, l = 1pt 
             [, tier=word2]
             [, tier=word2]
            ]
          ]
        ]
        [,tier=word]
      ]
      [
        [
          [,tier=word]
          [
            [,tier=word]
            [,tier=word]
          ]
        ]
        [, l = 50pt
          [,tier=word, draw, circle, fill=black
            [, edge=dotted, draw, circle, fill=black
              [, l = 12pt 
              	[,tier=word2]
              	[,tier=word2]
              ]
            ]
          ]
          [,tier=word]
        ]
      ]
    ]
  ]
 ]
]
\end{forest}}
\end{center}
\caption{Illustration of some day $i - 1$ (and the beginning of day $i$) of Lenski's LTEE with $4$ founder individuals (bullets), their  offspring trees within day $i-1$, and the  sampling from day $i-1$ to $i$ (dotted), for an average clone size of $5$. The second founder from the left at day $i - 1$ (and its offspring) is lost due to the sampling, and the second founder from the right at day $i$ carries a new beneficial mutation (indicated by the square). Reprinted from \cite[Figure~1]{EBAW-BGPW19}, \copyright~2019, with permission from Elsevier.
\label{EBAW-forest}
}
\end{figure}
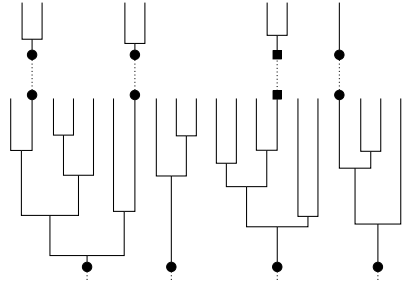

\index{Lenski's long-term evolution experiment}  
R.~Lenski started 12 replicates of  an experiment in 1988, and since then it has been running without interruption. This has become famous as the E.coli LTEE \cite{EBAW-Le19}. \index{LTEE}
Every morning,  a sample of $\approx\!5 \cdot 10^6$ \index{Escherichia coli} \emph{Escherichia~coli} \index{bacteria} bacteria is inserted into a defined amount of   fresh minimal glucose medium; let us call them the founder individuals (at day $i$, say). As soon as the nutrients are consumed, the bacteria  stop dividing  --- this is the case when the population has reached  a size of  $\approx 5 \cdot 10^8$ bacteria, with  $\approx 5 \cdot 10^6$ \index{clone} clones each of average size $\approx 100$, see Fig.~\ref{EBAW-forest}.   The next morning, the process is repeated by taking out of the $\approx 5 \cdot 10^8$ cells a random sample of $\approx 5 \cdot 10^6$, which then form founder individuals at day $i+1$. This induces a genealogy in discrete time $i=0,1,\ldots$ among the founder individuals.

The goal of the experiment is to observe
 evolution in real time. Indeed, the bacteria evolve via beneficial mutations, which
allow them to adapt to the environment and thus to reproduce faster.
One special feature of the LTEE \index{LTEE}  is that samples are frozen at regular intervals. They can be brought back to life at any time for the purpose of comparison  and thus form a living fossil record. In particular, one can, at any day~$i$, compare the current population with the initial (day-$0$) population via a  \index{competition experiment}  \emph{competition experiment} \cite{EBAW-LT94,EBAW-WRL13}, which yields the   empirical \index{fitness!Malthusian} (Malthusian) fitness at day $i$ relative to that at day 0; see below for a precise definition.
Fig.~\ref{EBAW-cannings_exp_init_0p15}  shows the time course over 21 years of the \index{fitness!relative} empirical relative fitness averaged over the  replicate populations,   as reported by \cite{EBAW-WRL13}. Obviously, the \emph{mean relative fitness} has a tendency to increase, but the increase levels off, which leads to a conspicuous concave shape.

In this section, we 
report on two closely related models
that describe the experiment by means of a \index{Cannings model} \emph{Cannings model} and explain the mean fitness curve. The first model was set up by  Gonz\'{a}lez-Casanova, Kurt, Wakolbinger, and Yuan \cite{EBAW-GKWY16} and will henceforth be referred to as the GKWY model; it assumes that (a) the \index{fitness!increment} fitness increments conveyed by beneficial mutations are deterministic and follow a regime of \index{epistasis!diminishing returns}\emph{diminishing returns epistasis} (that is, the increments decrease with increasing fitness), (b) a \index{weak mutation--moderate selection} weak mutation--moderate selection regime applies, and (c) the population is so large that a  \index{law of large numbers} law of large numbers is appropriate.  Building on this, the second model by Baake,   Gonz\'{a}lez-Casanova,  Probst, and Wakolbinger (\cite{EBAW-BGPW19}, referred to as the BGPW model) introduces additional random elements by (d) allowing for stochastic
fitness increments  and (e) 
considering effects of
\index{clonal interference} \emph{clonal interference}, which means 
that two mutants compete with each other for the success of going to \index{fixation} fixation. Both models build on earlier work by Wiser, Ribeck, and Lenski \cite{EBAW-WRL13}, who work close to the data and perform an approximate analysis in the spirit of theoretical biology, while we focus on  precise definitions of the models, on population-genetic concepts,  and on mathematical rigour where possible. Our presentation will be guided by~\cite{EBAW-BGPW19}.
\subsection{Interday and intraday dynamics}\label{EBAW-IDID}
Both the GKWY and the BGPW models take two distinct dynamics into account, namely, the dynamics \emph{within each individual day}, and the dynamics \emph{from day to day}. We will now explain these building blocks. Here, as well as in  Sections~\ref{EBAW-Heur} and \ref{EBAW-LenskiLLN}, we will focus on the case of deterministic fitness increments and the absence of clonal interference (as considererd in the GKWY model); the extensions in the BGPW model will be discussed  in Section~\ref{EBAW-BGPW}.

\smallskip

\paragraph{\textbf{Intraday dynamics}} 
Let $T$ be (continuous) physical time  within a day, with $T=0$ corresponding to the beginning of the  growth phase. Day $i$ starts with
$N$  founder individuals ($N \approx 5 \cdot 10^6$ in the experiment). The reproduction rate or \index{fitness!Malthusian} \emph{Malthusian fitness}  of founder individual~$j$ at day $i$ is 
$R_{ij}^{}$, where $i \geqslant 0$ and $1  \leqslant j \leqslant N$. It is assumed that at day 0 the population is \emph{homogeneous} (or \index{monomorphic} \emph{monomorphic}), that is,    $R_{0j}^{} \equiv R_0^{}$. Offspring inherit the reproduction rates from their parents.

We  denote by \index{rescaling}
\begin{equation}
	t = R_0 T	\quad \text{and } r^{}_{ij}  = \frac{R_{ij}}{R_0} \label{t} 
\end{equation}
dimensionless time and rates, so that on the time scale $t$ there is, on average, one split per time unit at the beginning of the experiment, so $r^{}_{0j} \equiv 1$.  We consider the $r_{ij}^{}$ as given (non-random) numbers.

We thus have  $N$ independent \index{Yule process} \emph{Yule processes}  at day $i$:  all descendants of founder individual $j$ (the members of the $j$-clone) branch at rate $r^{}_{ij}$, independently of each other. They do so until $t=\sigma_i$, where
$\sigma_i$ is the duration of the growth phase  on day $i$. We define $\sigma_i$ as the value of $t$ that satisfies
\begin{equation}\label{EBAW-def_sigma}
\EE(\text{population size at time } t) 
= \sum_{j=1}^N \ee^{r^{}_{ij}t} =\gamma N, 
\end{equation}
where $\gamma$ is, equivalently, the multiplication factor of the population within a day, the average clone size,  and the dilution factor from day to day in the experiment ($\gamma \approx 100$ in the LTEE). Note that, in the definition of $\sigma_i$, we have idealised by replacing the random population size  by its expectation. Since $N$ is very large, this is well justified, because the fluctuations of the random time needed to grow from size $N$ to size $100N$ in size are small relative to that time's  expectation. 

\smallskip

\paragraph{\textbf{Interday dynamics}} At the beginning of day $i > 0$, one samples $N$ new founder individuals out of the $\gamma N$ cells from the population at the end of day $i-1$. We assume that one of these new founders carries a \index{mutation!beneficial} \emph{beneficial mutation} with probability $\mu$; otherwise (with probability $1 - \mu$), there is no beneficial mutation. We think of $\mu$ as the probability that a beneficial mutation occurs in the course of day $i-1$ and  is sampled for day $i$.

Assume that the new beneficial mutation at day $i$ appears in individual $m$, and that the reproduction rate of the corresponding founder individual $k$ in the morning of day $i-1$ has been  $r_{i-1, k}$. The new mutant's reproduction rate is then assumed to be
\begin{equation}\label{EBAW-r_increase}
	r^{}_{im} = r^{}_{i-1,k} + \delta(r_{i-1,k}) \; \text{ with } \delta(r) \mathrel{\mathop:}= \frac{\varphi}{ r^{q}}.
\end{equation}
Here, $\varphi$ is the  beneficial effect due to the first mutation (that is $\delta(1)$),  and $q$ determines the strength  of \index{epistasis} epistasis. In particular, 
$q=0$ implies constant \index{fitness!increment} increments (that is, \index{fitness!additive} fitness is additive), whereas
$q>0$ means that the increment decreases with $r$, that is, we have \index{epistasis!diminishing returns} \emph{diminishing returns epistasis}. 
Note  that we only take into account beneficial mutations and adhere to the simplistic assumption that the \index{fitness!landscape} fitness landscape is \emph{permutation invariant}, that is, every beneficial mutation on the same background conveys the same deterministic fitness increment, no matter where it appears in the genome; this simplification is already used by Fisher in his  \index{staircase model} \emph{staircase model} \cite{EBAW-Fi1918} and is still common in the modern literature  \cite{EBAW-DeFi07}.   The assumption will be relaxed in Sec.~\ref{EBAW-BGPW}, where we turn to stochastic increments. 

\smallskip

\paragraph{\textbf{Mean relative fitness}} Let us now define the \index{fitness!mean relative} {\em mean relative fitness}, depending on the  reproduction rates $r_{ij}$ of the $N$ individuals in the sample at the beginning of day $i$, as
\begin{equation}\label{EBAW-rel_fitness}   
 F_i   \mathrel{\mathop:}=    \frac{1}{\sigma_i} \log \Big ( \frac{1}{N} \sum_{j=1}^{N} \ee^{r^{}_{ij}  \sigma_i^{}} \Big ).
\end{equation}
Here, $\sigma_i$ is as defined in~\eqref{EBAW-def_sigma}.
Note  that
\eqref{EBAW-rel_fitness} implies that
\begin{equation}\label{EBAW-exp_rel_fitness}
 \ee^{F_i^{} \sigma_i^{}}   =   \frac{1}{N} \sum_{j=1}^{N} \ee^{r^{}_{ij} \sigma_i^{}}.  
\end{equation}
Thus,  $F_i$ may be understood as the \index{effective reproduction rate} \emph{effective reproduction rate} of the population at day $i$, which  differs from the mean \index{fitness!Malthusian} Malthusian fitness $\frac{1}{N} \sum_j r^{}_{ij}$ unless the population is homogeneous.

\subsection{Heuristics for the power law of the mean fitness curve}\label{EBAW-Heur}
In a homogeneous population of relative fitness $F$, 
the length of the growth period  is 
\begin{equation}\label{EBAW-sigma}
\sigma(F) =\frac{\log \gamma}{F}
\end{equation}
(since this solves $\ee^{Ft}=\gamma$, cf. \eqref{EBAW-def_sigma}). 
It is crucial to note that \emph{the length $\sigma$ of the growth period decreases with increasing} $F$.

Assume a new mutation arrives in a homogeneous population of relative fitness $F$.  It conveys to the mutant individual a relative \index{fitness!increment} \emph{fitness increment} 
\begin{equation}\label{EBAW-delta_N}
\delta (F) = \frac{\varphi}{F^q},
\end{equation}
that is, the mutant has relative Malthusian fitness $F+\delta (F)$.
We define the \index{selective advantage} \emph{selective advantage} of the mutant as 
\begin{equation}\label{EBAW-s_N}
s(F) = \delta(F) \, \sigma(F).
\end{equation}
This is because the selective advantage is the product of the fitness increment and the duration  of a generation, which, in our case, is the time required for the population to grow to $\gamma$ times its original size, namely $\sigma(F)$; for details, see \cite{EBAW-Chevin11}, \cite[p.~1977]{EBAW-Sanjuan10}, and \cite{EBAW-BGPW19}.

It is essential to note that  $s$ in \eqref{EBAW-s_N} inherits the dependence on $F$ from $\sigma$ and thus $s$ decreases with increasing $F$ even for $q=0$. This is what we call the
\index{runtime effect}  \emph{runtime effect:} adding a constant to an interest rate $F$ of a savings account becomes less efficient when the runtime decreases.

Furthermore, it is precisely this notion of selective advantage conveyed by \eqref{EBAW-s_N}   that governs the \index{fixation!probability} \emph{fixation probability}. Namely, the fixation probability of the mutant turns out to be 
\begin{equation}\label{EBAW-pi_N}
\pi(F) \sim C\, s(F).
\end{equation}
Here, $\sim$ means asymptotic equality in the limit $N \to \infty$,\footnote{That is, $\pi(F) / (C \, s(F)) = \pi^{}_N(F) / (C \, s^{}_N(F)) \to 1$ as \mbox{$N \to \infty$}, in the setting of Section~\ref{EBAW-LenskiLLN}.} and $C \mathrel{\mathop:}= \gamma/(\gamma - 1)$. The key to understanding the role of $C$ is to formulate the interday dynamics in terms of a \index{Cannings model} \emph{Cannings model}.  In a neutral setting, this classical model of population genetics works by assigning in each time step  to each of $N$ (potential) mothers indexed $j=1,\ldots, N$ a random number $\nu_j$ of daughters   such that the  $\nu_j$ add up to $N$ and are \index{exchangeable} {\em exchangeable}, that is, they have a joint distribution that is invariant under permutations of the mother's indices \cite[Ch.~3.3]{EBAW-Ewens04}. In  \cite{EBAW-GKWY16}, the mothers are identified with the founders in a given day and the daughters with the founders in the next day, thus resulting in an extension of the \index{Cannings model}  Cannings model that includes mutation and selection. 
This extension is obtained by decreeing that a mutant with a selective advantage $s$ compared to the resident type has an expected offspring of size $(1 + s)$ times the expected offspring of a resident. (For the large class of Cannings models that admit a paintbox representation, a graphical construction that includes directional selection is given in \cite{EBAW-BGPoW19}.)

In our situation, the \index{offspring variance}  \emph{offspring variance} $v$ in one Cannings generation satisfies  
\begin{equation} 
\label{EBAW-ourv}
v = \mathbb{V}(\nu_1) \sim 2 \, \frac{\gamma-1}{\gamma} = \frac{2}{C}
\end{equation}
(\cite{EBAW-GKWY16}, see also \cite{EBAW-BGPW19} for an  explanation in terms of \index{pair coalescence probability} pair coalescence probabilities in the Cannings model).
 Hence \eqref{EBAW-pi_N} is in line with \index{Haldane's formula} \emph{Haldane's formula}
\begin{equation}\label{EBAW-Haldane}
\pi \sim \frac s{v/2},
\end{equation}
which relies on a \index{branching process!approximation} branching process approximation of the initial phase of the mutant growth; see \cite{EBAW-PW08} for an account of this method, including a historic overview.

Another crucial ingredient of the heuristics is the time window of length 
\begin{equation}\label{EBAW-u_N}
u(F) \sim \frac{\log \big (N \, s(F) \big )}{s(F)} 
\end{equation}
after the appearance of a beneficial mutation that will survive drift (a so-called \index{mutation!contending} \emph{contending mutation}); this results from a branching process approximation of the expected  time it takes for the mutation to become dominant in the population, see  \cite{EBAW-MS76,EBAW-DeFi07,EBAW-BGPW19}. Indeed, let $(Z_i)_{i \geqslant 0}$ be a Galton-Watson process with offspring mean $1+s$ and $s$ nonnegative and small. Then according to \eqref{EBAW-Haldane} we have the  asymptotics
\begin{equation}\label{EBAW-GWcond}
\mathbb E[Z_i \mid Z_i>0] \sim \frac v2 \frac{(1+s)^i}s.
\end{equation}
Hence, for any $\varepsilon > 0$, the expression in \eqref{EBAW-u_N} is asymptotically equal to that generation for which \eqref{EBAW-GWcond} reaches $\varepsilon N$.

\begin{figure}
 \begin{center}
    \includegraphics[width=5cm]{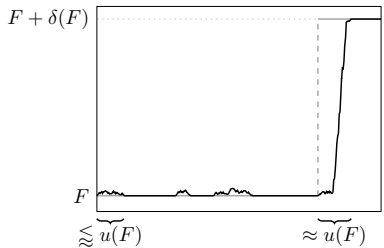}
\end{center}
\caption{\label{EBAW-heuristicLLN}
Schematic drawing of the relative fitness process (black) and the approximating jump process (grey). 
Reprinted from \cite{EBAW-BGPW19} Figure~3, \copyright~2019, with permission from Elsevier.
}
\end{figure}

All this now leads us to the dynamics of the relative fitness process.  As illustrated in Fig.~\ref{EBAW-heuristicLLN}, most mutants only grow to small frequencies and are then lost again (due to the sampling step). But if  a mutation does survive the initial fluctuations and gains appreciable frequency, then the dynamics turns into an asymptotically deterministic one and takes the mutation to \index{fixation} fixation quickly, cf.~\cite{EBAW-GL98,EBAW-DeFi07}, or \cite[Ch.~6.1.3]{EBAW-Du08}.  Indeed, within  time  $u(F)$, the mutation has either disappeared or gone close to fixation. Moreover, in the scaling regime \eqref{EBAW-scaling} specified in Section~\ref{EBAW-LenskiLLN}, this time is much shorter than the mean interarrival time $1/\mu$ between successive beneficial mutations.  As a consequence, there are, with high probablity, at most two types present in the population at any given time (namely, the \index{resident} \emph{resident} and the \index{mutant} \emph{mutant}), and \index{clonal interference} \emph{clonal interference is absent}. Therefore, in the scenario considered, survival of  drift  is equivalent to fixation. The parameter regime $u \ll 1/\mu$ is known as the \index{periodic selection} \emph{periodic selection} or \index{sequential fixation} \emph{sequential fixation} regime, and the resulting class of \index{origin-fixation model} \emph{origin-fixation} models is reviewed in \cite{EBAW-McCS14}. 

Next, we consider the expected per-day increase  in relative fitness, given the current value~$F$. This is
\begin{equation}\label{EBAW-E_F}
\EE(\Delta F \mid  F) \,  \approx \, \mu \, \pi(F) \, \delta (F) \\
 \sim \, \frac{\Gamma}{F^{2q+1}}.
\end{equation}
Here, the asymptotic equality is due to \eqref{EBAW-delta_N}--\eqref{EBAW-pi_N}, and the compound parameter
\begin{equation}\label{EBAW-compound}
\Gamma \mathrel{\mathop:}= C \, \mu \, \varphi^2 \log \gamma
\end{equation}
is the rate of fitness increase per day at day~0 (where $r_{0j}^{} \equiv F_0^{}= 1$). Note that $\varphi/F^q$ appears squared in the asymptotic equality in \eqref{EBAW-E_F} since it enters both $\pi$ and $\delta$. Note also that the additional $+1$ in the exponent of $F$ comes from the factor of $1/F$ in the length of the growth period \eqref{EBAW-sigma}, and thus reflects the \index{runtime effect} runtime effect.  The effect would be absent if, instead of our \index{Cannings model} Cannings model, a  discrete-generation scheme were used, as in \cite{EBAW-WRL13}; or a standard \index{Wright--Fisher model} Wright--Fisher model,  for which \cite{EBAW-KTP09}  calculated the expected fitness increase  and the fitness trajectory for various  \index{fitness!landscape} fitness landscapes, including the one in \eqref{EBAW-delta_N}. 

Eq. \eqref{EBAW-E_F} now leads us to  define  a new time variable $\tau$  via
\begin{equation}\label{EBAW-tau_Gamma}
i = \Big \lfloor \frac{\tau}{\Gamma} \Big \rfloor
\end{equation}
with $\Gamma$ of \eqref{EBAW-compound}; this means that one unit of time $\tau$ corresponds to $\Gamma$ days. With this \index{rescaling} rescaling of time, Eq. \eqref{EBAW-E_F}
corresponds to the \index{ordinary differential equation}  differential equation 
\begin{equation}\label{EBAW-ODE}
\frac{\dd{}}{\dd{\tau}} f(\tau) = \frac{1}{f^{2q+1}(\tau)}, \quad f(0)=1,
\end{equation}
with solution
\begin{equation}\label{EBAW-LLN}
  f(\tau) = \big ( 1+2\, (1+q) \, \tau \big )^{\frac{1}{2(1+q)}}.
\end{equation}
That is, the fitness trajectory follows a \index{power law} power law (and is concave).
Note that \eqref{EBAW-ODE} is just a \index{scaling limit} scaling limit of \eqref{EBAW-E_F}, where the expectation was  omitted due to a dynamical \index{law of large numbers} law of large numbers, as will be explained next.
\subsection{Scaling regime and \index{law of large numbers} law of large numbers}\label{EBAW-LenskiLLN}
We now think of $\mu = \mu^{}_N$ and $\varphi= \varphi^{}_N$ as being indexed with population size because the law of large numbers \index{law of large numbers} requires to consider a sequence of processes indexed with $N$. Thus, other   quantities now also depend on $N$ (so $\delta=\delta_N^{}, s=s^{}_N$, $\pi=\pi^{}_N$, $\Gamma=\Gamma_N$ etc.), and so does the  \index{relative fitness process}  \emph{relative fitness process}  $(F_i)_{i \geqslant 0} = (F^N_i)_{i \geqslant 0}^{}$ with $F_i$ of \eqref{EBAW-rel_fitness}. More precisely, we will take a \index{weak mutation--moderate selection} \emph{weak mutation--moderate selection limit}, which requires that $\mu^{}_N$ and $\varphi^{}_N$ become small in some controlled way as $N$ goes to infinity. Specifically, it is assumed in \cite{EBAW-GKWY16} that
\begin{equation}\label{EBAW-scaling}
\mu^{}_N \sim \frac{1}{N^{a}}, \; \varphi^{}_N \sim \frac{1}{N^{b}} \quad \text{as} \quad N \to \infty, \;
0 < b < \frac{1}{2}, \;  a > 3\,b.
\end{equation}
These assumptions will enter in Theorem \ref{EBAW-thm_LenskiLLN} below.
Due to the assumption $a> 3\,b$, $\mu^{}_N$ is of much lower order than $\varphi^{}_N$.  This is  used in  \cite{EBAW-GKWY16} to prove that, as $N\to \infty$,   with high probability  no more than two fitness classes are simultaneously present in the population over a long time span. Here is a quick intuitive reason for the bound $a>3b$ in \eqref{EBAW-scaling}. Reaching a macroscoping increase of the relative fitness requires asymptotically (as $N\to \infty$) no more than $\varphi_N^{-1-\varepsilon}$ successful mutations (with some $\varepsilon >0$), and in view of \eqref{EBAW-Haldane}, between two successful muations there are asymptotically no more than $\varphi_N^{-1-\varepsilon}$ unsuccessful ones. Because of \eqref{EBAW-u_N}, the time until a new mutation has either disappeared or gone to fixation can be estimated from above in probability by $\varphi_N^{-1-\varepsilon}$. The condition $a>3b$ ensures that the expected number of mutations that arrive in a total time of $(\varphi_N^{-1-\varepsilon})^3$ is asymptotically negligible. Indeed, as proved in \cite{EBAW-GKWY16}, this condition guarantees a strict absence of clonal interference as $N\to \infty$;  there it is conjectured that Theorem \ref{EBAW-thm_LenskiLLN}  holds even under the weaker condition $a>b$.
  
Furthermore, the scaling of $\varphi^{}_N$ implies that selection is stronger than genetic drift as soon as the mutant has reached an appreciable frequency. The method of proof applied in  \cite{EBAW-GKWY16} requires the assumptions \eqref{EBAW-scaling} in order to guarantee a \index{coupling} coupling between the new mutant's offspring and two nearly-critical \index{Galton--Watson process} Galton--Watson processes, between which the mutant offspring's size is `sandwiched' for sufficiently many days. Specifically, under the assumption $0 < b < \frac{1}{2}$,  the coupling  works  until the mutant offspring in our \index{Cannings model} Cannings model has reached a small    (but strictly positive) proportion of the population, or has disappeared. 
For the case $\frac{1}{2} < b < 1$, \index{Haldane's formula} Haldane's formula \eqref{EBAW-Haldane}  holds  for a very closely related class of Cannings models \cite[Theorem 3.5]{EBAW-BGPoW19};  the proof relies on \index{duality} duality methods invoking an \index{ancestral selection graph} ancestral selection graph in discrete time.
We  conjecture that  the assertion of Theorem \ref{EBAW-thm_LenskiLLN} holds also for $0 < b < 1$.

In the case where selection is much stronger than mutation, the classical models of population genetics, such as the \index{Wright--Fisher model} Wright--Fisher or \index{Moran model} Moran model, display the well-known dynamics of \index{sequential fixation} sequential fixation \cite{EBAW-McCS14}, that is,   a new beneficial mutation is either lost quickly  or goes to fixation. Qualitatively, our Cannings model displays a  similar behaviour.
Furthermore, as already indicated, with the chosen \index{scaling}  scaling the population turns out to be homogeneous on generic days~$i$ as $N\to \infty$. This has the following  practical consequence for the relative fitness process $(F_i^N)_{i \geqslant 0}^{}$.
On a time scale with a unit of $1/(\mu^{}_N \, \varphi^{}_N)$ days, $(F^N_i)_{i \geqslant 0}^{}$ turns into a jump process as $N \to \infty$, cf. Fig.~\ref{EBAW-heuristicLLN}. 
%
The precise formulation of the limit law \cite{EBAW-GKWY16} reads as \index{weak mutation--moderate selection}

\begin{theorem}\label{EBAW-thm_LenskiLLN}
For $N \to \infty$ and under the  scaling \eqref{EBAW-scaling}, the sequence of processes $\big (F^N_{\lfloor \tau /\Gamma_N \rfloor} \big )_{\tau \geqslant 0}$ converges,
in distribution and locally uniformly, to the deterministic function $\big ( f(\tau) \big )_{\tau \geqslant 0}$ in \eqref{EBAW-LLN}.
\end{theorem}

The theorem was proved  along the heuristics outlined above\footnote{Note that \cite{EBAW-GKWY16} partly works with dimensioned variables, which is why the notation and the result look somewhat different.}. 
It is a \index{law of large numbers} reasoning that allows to go from \eqref{EBAW-E_F} to \eqref{EBAW-ODE} (and thus to `sweep the expectation under the carpet'), in the following sense. For large $N$ and under the scaling assumption \eqref{EBAW-scaling}, fitness is the sum of a large number of small per-day \index{fitness!increment}  increments accumulated over many days, and may be approximated by its expectation. 

Since time has been rescaled via \eqref{EBAW-tau_Gamma}, Eq.~\eqref{EBAW-LLN} has $q$ as its single parameter. Note that  $1/(2\,(1+q))<1$ (leading to a concave $f$)  whenever $q \geqslant 0$; in particular, \emph{the fitness curve is concave even for} $q=0$, \emph{that is, in the absence  of  \index{epistasis} epistasis}. 
In contrast, the \index{fitness!trajectory} fitness trajectory obtained in \cite{EBAW-KTP09}  for the \index{Wright--Fisher model}  Wright--Fisher model under $q = 0$ is linear. The difference is due to the \index{runtime effect} runtime effect, which is present in our Cannings model even for $q=0$ because of the parametrisation of the intraday dynamics with the individual reproduction rate $r$: If the population as a whole already reproduces faster, then the end of the growth phase is reached sooner and thus leaves less time for a mutant to play out its advantage $\delta(r) = \varphi/r^0 = \varphi$ of \eqref{EBAW-r_increase}. The \index{Wright--Fisher model} Wright--Fisher model of \cite{EBAW-KTP09}  does not display the runtime effect because it does not contain the individual (intraday) reproduction rate as a parameter.
The second parameter, namely  $\Gamma_N$, reappears when  $\tau$ is translated back into days; that is,  $F^N_i \approx f(\Gamma_N \, i)$.

\subsection{Random fitness increments and clonal interference}
\label{EBAW-BGPW}
Let us now turn to \index{fitness!increment} random beneficial effects. To this end, we scale the fitness increments with a positive random variable $X$ with density $h$ and expectation $\EE(X)=1$. We assume throughout that  $0<  \EE(X^2) < \infty$ (the degenerate case $X \equiv 1$ requires special treatment, as detailed in \cite{EBAW-BGPW19}).

Taking into account the dependence on $X$, the quantities in \eqref{EBAW-delta_N}--\eqref{EBAW-pi_N} and \eqref{EBAW-u_N} 
turn into
\begin{equation}\label{EBAW-X}
\begin{split}
& \delta (F,X)  \, = \, X \, \frac{\varphi}{F^q}, \quad
\sigma(F)  \, = \, \frac{\log \gamma}{F} \text{ (as before)}, \quad
s(F,X) \,  = \, \delta(F,X) \, \sigma(F), \\
& \pi(F,X) \,  \approx \, C \, s(F,X), \quad
u (F,X) \,  = \, \frac{\log (N\,s(F,X))}{s(F,X)} \approx \frac{\log (N\, \varphi\, X)}{s(F,X)}
\end{split}
\end{equation}
  (see\ \cite{EBAW-BGPW19} for an explanation of the approximation for $u$.) Here we use $\approx$ in place of $\sim$ because --- in contrast to Sec.\ref{EBAW-LenskiLLN} --- we have no limit law available so far.
Note  that large $X$ implies large $s$ and hence small $u$ and vice versa.
The following Poisson picture will be central to our heuristics: %
The process of \index{mutation!beneficial} \emph{beneficial mutations} with scaled effect $x$ that arrive at time $\tau$ has intensity $\mu \dd{\tau} h(x) \dd{x}$ with points $(\tau, x) \in \mathbb R_+\times \mathbb R_+$. %
And in fitness background $\approx F$, we denote by $\Pi$ the \index{Poisson process} Poisson process of \index{contending mutation} \emph{contending mutations}, which has intensity $\mu \dd{\tau} h(x) \, \pi(F,x) \dd{x}$ on $\mathbb R_+\times \mathbb R_+$. Recall that contending mutations are those  that survive drift;
but since we now take into account clonal interference, contending mutations do not necessarily go to \index{fixation} fixation.%

We now describe a refined version of the \index{Gerrish--Lenski heuristics} \emph{Gerrish--Lenski heuristics} for \index{clonal interference} \emph{clonal interference}, adapted to the context of our model.
Verbally, the heuristics says that `if two contending mutations appear within the time required to become dominant in the population, then the fitter one wins.' We therefore posit that
if, in the fitness background $\approx F$, two contending mutations $(\tau, x)$ and $(\tau', x')$ appear at $\tau < \tau' < \tau + u(F,x)$, then the first one outcompetes (`kills') the second  if $x' \leqslant x$, and the second one kills the first  if $x' > x$. Thus, neglecting interactions of higher order, given that a contending mutation arrives at $(\tau, x)$ in the fitness background $\approx F$, the probability that it does not encounter a killer in its  past is %
\begin{equation}
\overleftarrow{\chi}(F, x) \mathrel{\mathop:}= 
 \exp \Big ( -  \int_x^\infty \!\!\!  \mu \, \pi (F,y)\,  u(F,y) \, h(y) \dd {y} \Big ), \label{EBAW-past}
\end{equation}
whereas the probability that it does not encounter a killer in its future is 
\begin{equation}
 \overrightarrow{\chi}(F, x) \mathrel{\mathop:}= 
\exp \Big ( - u (F, x) \int_x^\infty \!\!\!  \mu \, \pi (F,x') \, h(x') \dd{x'} \Big )
\label{EBAW-future}
\end{equation}
(note that only the term corresponding to $\overrightarrow{\chi}$ is considered by \cite{EBAW-GL98}; the inclusion of $\overleftarrow{\chi}$ makes the heuristics consistent with its verbal description).
Using \eqref{EBAW-X}, $\overleftarrow{\chi}(F, x)$ is approximated by
\begin{align}
\overleftarrow{\psi}(x) & \mathrel{\mathop:}= 
\exp  \Big ( -\mu\, C   \int_x^\infty  \log (N \, \varphi \, y) h(y) \dd{y}  \Big ), \label{EBAW-psipast}
\intertext{whereas $\overrightarrow{\chi}(F, x)$ is approximated by}
\overrightarrow{\psi}(x) & \mathrel{\mathop:}= 
 \exp  \Big ( -\mu  \, \frac{C \log (N \, \varphi \, x)}{x} \int_x^\infty x' \,  h(x') \dd{x'}  \Big ).
\label{EBAW-psifuture}
\end{align}
Note that neither $\overleftarrow{\psi}$ nor $\overrightarrow{\psi}$ depend on $F$. %
Thus, setting $\overleftrightarrow{\chi} \mathrel{\mathop:}= \overleftarrow{\chi} \, \overrightarrow{\chi}$ and analogously $\overleftrightarrow{\psi}\mathrel{\mathop:}= \overleftarrow{\psi} \, \overrightarrow{\psi}$, we obtain, as an analogue of~\eqref{EBAW-E_F}, the expected (per-day) increase of $F$, given the current value of $F$, as
\begin{equation} 
 \EE  (\Delta F \mid  F) 
  \approx \,  \mu \int_0^\infty \!\!\! \delta (F,x)  \, \pi(F,x) \,
\overleftrightarrow{\chi}(F, x) \, h(x) \dd{x}   
 \approx \,    \frac{\Gamma}{F^{2q+1}},   \label{EBAW-exp_incr_random}  
 \end{equation}
where
\begin{equation}
  \Gamma \mathrel{\mathop:}= C \, \mu \, \varphi^2 \log (\gamma) \, I(\mu, \varphi) 
  \label{EBAW-compound_stoch}
\end{equation}
and $I(\mu, \varphi) \mathrel{\mathop:}= \EE \big (\overleftrightarrow{\psi}(X) \, X^2 \big)$. %
Similarly as in Sec.~\ref{EBAW-Heur}, the assumption of a suitable \index{concentration} concentration of the random variable $\Delta F$ around its conditional expectation allows us to take \eqref{EBAW-exp_incr_random} into
\[
F_{\lfloor   \tau / \Gamma \rfloor} \approx \EE \big (F_{\lfloor   \tau / \Gamma \rfloor}  \big ) \approx f(\tau) 
\]
with $f$ as in \eqref{EBAW-LLN}. %
This 
means that an approximate
\index{power law} \index{fitness!trajectory} \emph{power law of the mean fitness curve}  \emph{applies under any suitable distribution of fitness effects}; in particular, the \emph{epistasis parameter} $q$ is \emph{not affected by the  distribution} of $X$. %

\begin{figure}\centering
\resizebox{\textwidth}{!}{\input{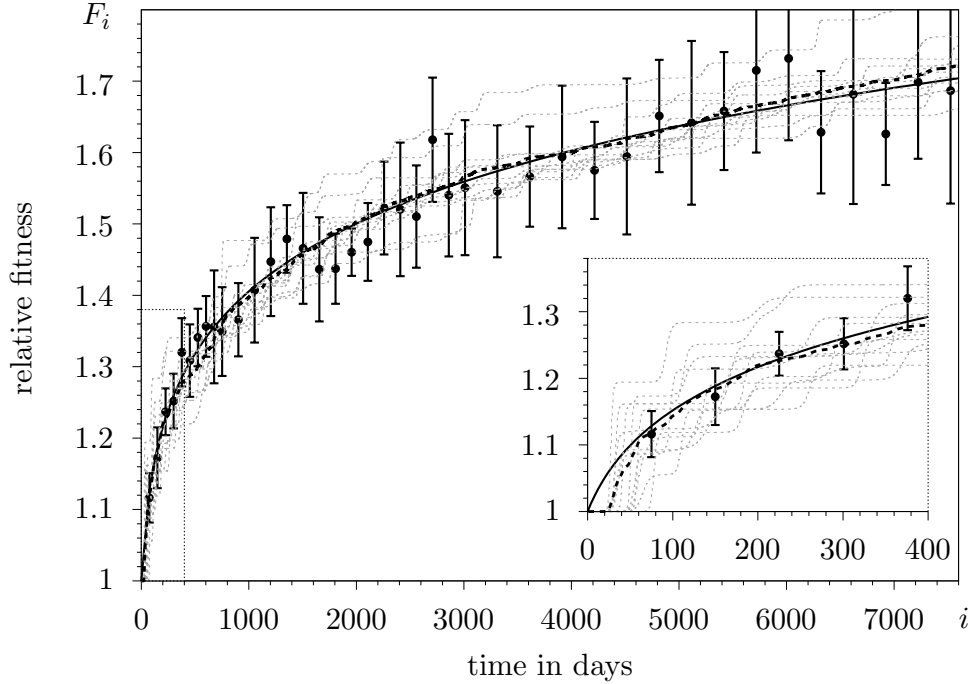}}
\caption{The fitness curve in model and experiment. Bullets: empirical relative fitness averaged over all 12 populations  with error bars (95\% confidence limits based on the 12 populations) from \cite[Fig. 2A and Table S4]{EBAW-WRL13};  solid  line: $F^{}_i \approx f(\Gamma \, i)$ with $f$ of \eqref{EBAW-LLN} and  parameter values $q=4.2$  and $\Gamma=3.2 \cdot 10^{-3}$ obtained by a least-squares fit  to the data; grey dashed lines: 12 individual trajectories $F_i$ obtained via  simulations of the Cannings model with $X$ following Exp(1) and parameters  $N=5 \cdot 10^6, \gamma=100, \varphi = 0.0375$, and $\mu=0.73$ (the values for $\mu$ and $\varphi$ are obtained from the fitted value of $\Gamma$ via \eqref{EBAW-compound_stoch} combined with the  value of the mean
fitness increment of the first fixed beneficial mutation as observed in independent experiments \cite{Lenski91}); black dashed line: average over the 12 simulations; inset: zoom on the early phase.
\label{EBAW-cannings_exp_init_0p15}
}
\end{figure}

\paragraph{\textbf{Exponentially distributed beneficial effects}} %
For definiteness, we now specify $X$ as following Exp$(1)$, the exponential distribution with parameter~1. %
This was the canonical choice also in previous investigations (cf.\ \cite{EBAW-GL98,EBAW-WRL13}) and is in line with experimental evidence (reviewed by \cite{EBAW-EWK07}) and  theoretical predictions \cite{EBAW-Gill84,EBAW-Orr03}. %
Fig.~\ref{EBAW-cannings_exp_init_0p15} shows the corresponding  simulations (the estimation of the parameters $\mu$ and $\varphi$ is an art of its own, for which we refer to \cite{EBAW-BGPW19}). The  simulation mean agrees nearly perfectly with the approximating power law.  However, for $i$ small, the fluctuations in the simulation are larger than those observed, whereas for large $i$, this is reversed. While we have no convincing explanation for the former, the latter may   be due to the constant parameters assumed by the model, whereas parameters do vary across replicate populations in the experiment.

\section[A host-parasite model with  balancing selection // and reinfection \phantom{AAA}]{A host-parasite model with  balancing selection \\ and reinfection}
\index{host-parasite model} \index{selection!balancing}
In this section, we review  a model motivated by observations concerning the
\index{human cytomegalovirus} human cytomegalovirus  \index{HCMV} (HCMV), an old herpesvirus, which is carried by a substantial
fraction of mankind \cite{EBAW-Ca10}. 
In DNA data of HCMV, a high genetic diversity is observed in coding regions, see \cite{EBAW-PuGoe11}. This diversity can be helpful to resist the defense of the host.
Furthermore, for guaranteeing its long-term survival, HCMV seems to have developed elaborate mechanisms
that allow it to
persist lifelong in its host and to establish \index{reinfection} reinfections in already infected hosts. The purpose of the model described in the sequel is to study the effects
of these mechanisms on the maintenance of diversity
in a parasite population. A central issue hereby is that, due to the \index{reinfection} reinfections, the diversity of the (surrounding) parasite population can be transmitted to the individual hosts. Our presentation will be guided by \cite{EBAW-PoWa19+}.
 \subsection{A hierarchical Moran model}
 \index{Moran model} \index{host-parasite model}
Consider a population consisting of $M$ hosts, each of which carries a population  of $N$ parasites. For simplicity we assume that $M$ and $N$ remain constant over time, and that there are two types of parasites, $A$ and $B$. We model the evolution of the parasite
population distributed over the hosts  by a $\{0,\tfrac1N,\ldots,1 \}^M$-valued Markovian jump process
${\bf X}^{N,M}= (X^{N,M}_1(t), ..., X^{N,M}_M(t))_{t \geq 0}$, where $X^{N,M}_i(t)$ and $1-X^{N,M}_i(t)$, $1\le i\le M$,  represent the relative frequencies
of type $A$- and type $B$-parasites, respectively, in host $i$ at time~$t$.
Before stating the jump rates of ${\bf X}^{N,M}$ in Definition \ref{EBAW-jumpratesMN}, we describe the dynamics  in words. The host population as well as the parasite population within each host follow dynamics that are modifications of the classical
Moran dynamics. We work at the time scale of \index{host replacement}  host replacement, that is, every host dies at rate 1 and is replaced by an offspring of a uniformly chosen member of the host population. (See Remark \ref{EBAW-RfiniteM} (f) for a possible generalisation of the latter assumption.) The $N$ parasites of the new host all carry  the type of a parasite chosen randomly from the parasite population of its parent. On this time scale, the
reproduction rate of each parasite is assumed to be $g_N$. The parasite population within a host experiences 
\index{selection!balancing} balancing selection towards an equilibrium frequency $\eta$ for some fixed $\eta \in (0,1)$.  More specifically, in host $i$ parasites of type $A$, when at relative frequency $x_i$, reproduce at rate
$g_N(1+s_N(\eta - x_i))$ and those of type $B$ at rate $g_N(1-s_N(\eta-x_i))$, where $s_N$ is a small positive number. 
At a reproduction event, a parasite splits into two and replaces a randomly chosen parasite from the same host. 
\index{reinfection} Reinfection events occur at rate $r^{}_N$ per host; then a single parasite in the reinfecting host (both of which are randomly chosen) is copied and transmitted to the reinfected
host. At the same time,  a randomly chosen parasite is instantly removed from this host;  this way the parasite population size in each of the hosts is kept constant.
This hierarchical host-parasite dynamics is summarised in the following
\allowdisplaybreaks
\begin{definition} \label{EBAW-jumpratesMN} In the process ${\bf X}^{N,M}$, 
jumps from state 
$\textbf{x} = (x_1, ..., x_M)  \in \{0,\tfrac1N,\ldots,1 \}^M$ occur for $i=1, ..., M$ 
\begin{align*}
&\mbox{to }  \textbf{x} + \frac{1}{N} e^{}_i  &  \mbox {at rate }  &  g^{}_N  (1+  s^{}_N (\eta- x^{}_i)) N x^{}_i (1- x^{}_i) +  r^{}_N   \frac{1}{M} \sum_{j=1}^M x^{}_j (1-x^{}_i)  \notag
  \\  
&\mbox{to } \textbf{x} - \frac{1}{N} e^{}_i  &  \mbox {at rate }  & g^{}_N  (1+ s^{}_N ( x^{}_i-\eta)) N x^{}_i (1- x^{}_i) + r^{}_N \frac{1}{M} \sum_{j =1}^M (1-x^{}_j) x_i   \label{EBAW-ViralModel} \\
&\mbox{to } \textbf{x} + (1- x^{}_i)e^{}_i &  \mbox {at rate }   & \bar{ {\bf x}}:= \sum_{j=1}^M x^{}_j , \qquad \mbox{ and }
\mbox{to } \textbf{x} - x^{}_i e^{}_i \quad   \mbox {at rate }    (1- \bar{ \bf x} ), \notag
\end{align*}
with  
$e_i=(0, ..., 1, ..., 0)$
the $i$-th unit vector of length $M$. 
\end{definition}
This scenario can be interpreted in classical population genetics terms as a population distributed over $M$ \index{island model} islands and \index{migration} migration between islands.
Within each island, reproduction is panmictic and driven by \index{selection!balancing} balancing selection. The \index{hierarchical model} model is {\em hierarchical} in the sense that also the population of hosts is evolving.

 Related hierarchical models have been studied from a mathematical perspective 
in \cite{EBAW-Da18} and \cite{EBAW-LuMa17}, for example. In these papers, an emphasis is on models for selection on two scales, and phase transitions
(in the mean-field limit) are studied, in which
 particularly the higher level of selection (namely group selection) can drive the evolution of the population.
In our model, \index{selection!balancing} balancing selection only acts at the lower level (i.e. the within-host parasite populations); but we focus on parameter regimes in which balancing
selection is also lifted to the higher level, 
such that both parasite types are maintained for a long time  in the host population that consists of hosts carrying a single parasite type only, as well as hosts carrying both types 
of parasites. This  corresponds to observations in samples of HCMV hosts, see 
 \cite{EBAW-PoGo19+} and the references therein. 
\subsection{Laws of large numbers and propagation of chaos}
\index{law of large numbers} \index{propagation of chaos}
In what follows, we  specify the assumptions on the strength of selection,  intensity of reinfection, and parasite reproduction relative to the rate of host replacement.
For a moderate strength of selection 
 we show 
that, in the limit of an infinitely large parasite population per host, only three 
states of typical hosts exist, namely those infected with only one of the types $A$ or $B$ and those infected with both types, where  $A$ 
is at frequency $\eta.$ These three host states will be called the {\it pure states} (if the frequency of type $A$ in a host is 0 or 1)
and the {\it mixed state} (if the frequency of type $A$ in a host is $\eta$). The selection is assumed to be weak enough that  many reinfection attempts are required until an effective reinfection occurs, and at the same time to be strong enough that an effective reinfection leads the parasite type frequency in the reinfected host  to the equilibrium frequency $\eta$ quickly.

Only \index{reinfection} reinfection events can change a host state from  pure to  mixed. In most cases 
reinfection is not effective, in the sense that it only causes  a short excursion from the boundary frequencies 0 and 1.
We will see that if the selection strength and reinfection rate are \index{scaling} appropriately scaled, the \emph{effective} reinfection  acts on the same time
scale as host replacement. Furthermore, if selection is of moderate strength and parasite reproduction is fast enough (but not too fast), 
transitions of the boundary frequencies to the equilibrium frequency~$\eta$ will appear as jumps on the host time scale, and transitions between
the host states 0, $\eta$ and 1 are only caused by \index{host replacement} host replacement and effective reinfection events.
It turns out (see Sec.~\ref{EBAW-poly}) that, if the effective \index{reinfection} reinfection rate is larger than a certain bound depending on $\eta$, then, 
 in the limit $N\rightarrow \infty$ and $M\rightarrow \infty$, there exists a stable equilibrium of the relative frequencies of hosts of types 0, $\eta$ and 1, 
at which both types of parasites are present in the overall parasite population at  non-trivial frequencies.

We now collect the assumptions  on the parameter regime, namely moderate \index{selection!balancing} selection in $(\mathcal{ A} 1)$,  frequent \index{reinfection} reinfection in $(\mathcal{ A} 2)$, and upper and lower bounds for a  fast parasite reproduction in $(\mathcal{ A} 3)$.  \\\\
\index{weak migration--moderate selection}
\textbf{Assumptions ($\mathcal{A}$)}: There exist  $b\in (0,1)$, $r>0$ and $\epsilon >0$ such that the parameters $s^{}_N$, $r^{}_N$ and $g^{}_N$ of Definition \ref{EBAW-jumpratesMN} obey
 \begin{itemize}
 \item[$(\mathcal{ A} 1)$]  \qquad \qquad\qquad\qquad\qquad $s^{}_N = \frac{1}{N^{b}},$ \\
 \item[$(\mathcal{ A} 2)$]    \qquad \qquad\qquad\quad\quad $\displaystyle \lim_{N\rightarrow \infty} r^{}_N s^{}_N = r, $\\
 \item[$(\mathcal{ A} 3)$]   \qquad \qquad $\displaystyle   \frac{1}{g^{}_N} = o\Big ( \frac{1}{N^{3b +\epsilon}} \Big ), \qquad 
 g^{}_N = \mathcal O \big (\exp(N^{1 - b (1+\epsilon)}) \big ).$
  \end{itemize}

%

\begin{remark}\label{EBAW-Rem2_2}
\index{scaling}
 Assumption $(\mathcal{ A} 1)$  implies that 
\[\lim_{N\rightarrow \infty} s^{}_N =0 \mbox{ and }
\lim_{N\rightarrow \infty} s^{}_N N = \infty,\] while assumptions $(\mathcal{ A} 1)$, $(\mathcal{ A} 2)$ $(\mathcal{ A} 3)$ together imply that for large $N$
\[1 \ll r^{}_N \ll g^{}_N. \]
 The latter says that hosts experience frequent \index{reinfection} reinfections during their lifetime, and  many parasite reproduction events happen between two reinfection events.  Such a parameter regime seems realistic;
 see \cite{EBAW-PoGo19+} for additional discussion.
\end{remark}

In what follows, we will analyse the cases `$N\to \infty$ with $M$ fixed' (Theorem \ref{EBAW-TfiniteM}), `first $N\to\infty$, then $M\to\infty$' 
(Proposition \ref{EBAW-chaos1}), and `$N\to\infty$, $M\to\infty$ jointly' (Theorem \ref{EBAW-MF}).

\smallskip

\paragraph{\bf Large parasite population, finite host population}\label{EBAW-Sec2.2} 
Let the number $M$ of hosts be fixed. We specify the jump rates of the $\{0,1,\eta\}^M$-valued Markovian jump process
 $\textbf{Y}^M=(Y^M_1(t), ..., Y^M_M(t))_{t\geq 0}$, which will turn out to be the process of type-$A$ frequencies
 in hosts $1, \dots, M$ in the limit $N\rightarrow \infty$. From  state 
 $\mathbf{y}= (y^{}_1, ...., y^{}_M)$, the process $\textbf{Y}^M$ jumps by flipping, for $i\in \{1,\ldots,M\}$, the component $y_i$ 
 \begin{alignat}{3}
& \text{\quad from \quad } 0 \text{ or } \eta \text{\quad to \quad }1 && \text{\quad at rate \quad }   \frac{1}{M} \sum_{j=1}^M y_j, \nonumber \\ 
&\text{\quad from \quad } 1 \text{ or } \eta \text{\quad to \quad }0 && \text{\quad at rate \quad }  \frac{1}{M} \sum_{j=1}^M (1- y_j), \label{EBAW-Yrates} \\
& \text{\quad from \quad } 0 \qquad \text{\quad to \quad }\eta && \text{\quad at rate \quad }\frac{ 2r \eta }{M} \sum_{j =1}^M y_j, \nonumber \\
& \text{\quad from \quad } 1 \qquad \text{\quad to \quad }\eta && \text{\quad at rate \quad } \frac{ 2r (1-\eta)}{M} \sum_{j =1}^M (1-y_j).\nonumber
\end{alignat}
The first two lines in the display \eqref{EBAW-Yrates} corrspond to host replacement, and the last two lines capture the effective  reinfection. That is, the coefficient $2r\eta$ in the third line should, in the light of assumption $(\mathcal A 2)$, be understood as the limit of the product of the  reinfection rate $r_N$ and the asymptotic \index{probability to balance} `probability to balance' $2s_N\eta$, see the explanation in Remark \ref{EBAW-RfiniteM} (c) below.
\begin{theorem}[{\cite[Theorem 1]{EBAW-PoWa19+}}]\label{EBAW-TfiniteM}
Let $\mathbf X^{N,M}$ be the $\{0,\frac 1N,\ldots,1\}^M$-valued process with jump rates  given in Definition \ref{EBAW-jumpratesMN}. Fix 
$M \in \mathbbm{N}$ and assume that the law of
 $\mathbf X^{N,M}(0)$ converges weakly,  as $N\to \infty$, to a distribution $\rho$ concentrated on  $(\{0\}\cup [\alpha,1-\alpha] \cup \{1\})^M$ for some $\alpha>0$. Let $\textbf{Y}^{\ts M}$  be the process with jump rates \eqref{EBAW-Yrates}, and
 with the  distribution of $\textbf{Y}^{\ts M}(0)$ being the image of~$\rho$  under the mapping 
 $0 \mapsto 0$, $1 \mapsto 1$, and $[\alpha, 1-\alpha] \ni x \mapsto \eta$.
Assume ($\mathcal A$) and let  $0 <\underline t <\overline t<\infty$. On the time interval~$[\underline t,\overline t]$ the sequence of processes
 $\textbf{X}^{N,M}$, $N=1,2,\ldots$, then converges, as $N\to \infty$,   to the process $\textbf{Y}^{\ts M}$,  in distribution with respect to the \index{Skorokhod topology} Skorokhod $\mathrm M_1$-topology (see Remark \ref{EBAW-RfiniteM} (e) for an intuitive description of this topology).
 \end{theorem}
 
\begin{remark}\label{EBAW-RfiniteM} (a) 
Theorem \ref{EBAW-TfiniteM} reveals the emergence of processes of jumps from the boundary points $0$ and $1$ to the equilibrium frequency $\eta$; these jumps capture the outcomes of effective reinfections. 

(b) The jump processes  $Y_i^M$ that arise as the \index{law of large numbers} limits of the sequences $X_i^{N,M}$ as $N\to \infty$ bear similarities to the jump process addressed before Theorem \ref{EBAW-thm_LenskiLLN}:  a jump from type $0$ (or type $1$) to type $\eta$ in host~$i$ is caused by a `successful reinfection' (with a parasite of the complementary type),  whereas in the context of the LTEE model, it is the `successful mutations' that count for the jumps of the mean fitness.

(c)  Essential quantities required to show the \index{concentration} concentration on the two pure frequencies and the mixed equilibrium  are the \index{probability to balance} {\it probability to balance}, i.e.
the probability with which a \index{reinfection} reinfection event leads to 
the establishment of the second type in a  host so far of pure type; and the \index{time to balance} {\it time to balance}, i.e.~the
time needed to reach (a small neighborhood 
of) the equilibrium frequency $\eta$ after reinfection. These quantities determine the parameter regimes
in which we can observe the described scenario. Similar to the case of directional selection (see e.g.~\cite{EBAW-Cha06, EBAW-PoPf13} and Section \ref{EBAW-sec1}), \index{branching process!approximation}
branching process approximations 
as well as approximations by (deterministic) ordinary differential equations 
can be used to estimate these probabilities and times. 
A~notable difference compared to the situation
of \index{selection!directional} directional selection described in Section \ref{EBAW-sec1}
 is a change of the (role of the) coefficient of selection $s$ in \index{Haldane's formula} Haldane's formula \eqref{EBAW-Haldane} for the \index{fixation!probability} fixation probability: according to the jump rates specified in Definition~\ref{EBAW-jumpratesMN}, when starting  from frequency $\frac 1N $, the coefficient of selection is $\sim s_N\eta$, while the offspring variance is again asymptotically equal to~1. Thus Haldane's formula predicts that the    probability to  `take off', and hence the probability to balance
(when starting from frequency $\frac 1N $), is asymptotically equal to $2s^{}_N\eta$ as  $N\to \infty$. This is proved in  \cite[Lemma 3.6]{EBAW-PoWa19+}.  
Furthermore the \index{time to balance} time to balance \cite[Proposition 3.8]{EBAW-PoWa19+} is
longer than the \index{fixation!time} fixation time in the corresponding setting of Section \ref{EBAW-sec1}.
This is due to the fact that random fluctuations close to the equilibrium are larger than fluctuations
close to the boundary. 

(d) The reason for considering,  in Theorem \ref{EBAW-TfiniteM}, time intervals $[\underline t,\overline t]$ instead of 
$[0,\overline t]$ is to disregard the jumps at time $0$ in the limiting process that result from the asymptotically instantaneous (as $N\to \infty$) stabilisation of the type frequencies due to the  \index{selection!balancing} balancing selection.

(e) After an effective \index{reinfection} reinfection of host $i$, the $i$-th component of $\mathbf X^{N,M}$ performs (when $N$ is large) a quick transition with small jumps starting from the boundary and leading close to $\eta$.   Thus the \index{Skorokhod topology} Skorokhod $\mathrm M_1$-topology, which is coarser than the more common \mbox{$\mathrm J_1$-topology},  adequately describes the mode of convergence of $\mathbf X^{N,M}$   to the jump process $\mathbf Y^M$ as $N\to \infty$.  For a definition and characterisation of these topologies see \cite{EBAW-Skorokhod56}. Roughly stated, two paths are close to each other in the $\mathrm J_1$-topology if they are uniformly close after a small (possibly inhomogeneous) time shift of one of them. For a similar notion in  the $\mathrm M_1$-topology, define the graph of a path by completing the path's jumps through vertical interpolations, and require that there exist parametrisations of the two graphs that are uniformly close to each other in space-time. Beside the  convegence (in the sense of finite-dimensional distributions) of the genealogies of $\textbf{X}^{N,M}$ towards that of  $\textbf{Y}^{M}$, which relies on elements of the graphical construction described in the next paragraph, the \index{tightness} tightness criterion for the  $\mathrm M_1$-topology provided by \cite[Theorem 3.2.1]{EBAW-Skorokhod56} is used in the proof of Theorem~\ref{EBAW-TfiniteM}.

(f) Stimulated by a referee's question, we conjecture that the essentials of the results of \cite{EBAW-PoWa19+}, which are reviewed here, remain valid if the hosts' lifetimes are generalised from i.i.d.\ standard exponentials to i.i.d.\ random variables with expectation 1. This would make the flip rates in the last line of Definition \ref{EBAW-jumpratesMN} and in the  first two lines  of \eqref{EBAW-Yrates} dependent also on the age of the respective host. Furthermore, it would introduce a renewal component into the dynamics of the process $V$ specified in Definition \ref{EBAW-defV},  but otherwise leave the statements of Theorem \ref{EBAW-TfiniteM}, Proposition \ref{EBAW-chaos1}, and Theorem \ref{EBAW-MF}  unchanged.

 
\end{remark}
Theorem \ref{EBAW-TfiniteM} assumes a finite  host population (of constant size), with each host carrying a large number of parasites.
However, in view of the discussion in~\cite{EBAW-PoGo19+}, it is
realistic to assume that  the number of infected hosts is also large. We will consider two cases: In the next paragraph we let first $N\rightarrow \infty$ and then $M\rightarrow \infty$; thereafter  we assume 
a joint convergence of $N$ and $M=M_N$ to $\infty$.

\smallskip

\paragraph{\bf Iterative limits: Huge parasite population, large host population}\label{EBAW-secMindepN}
 According to Theorem \ref{EBAW-TfiniteM}, the process $\textbf{Y}^M$  arises in  the limit of $N\to \infty$ parasites  per host, with the number $M$ of hosts  fixed.  $\textbf{Y}^M$ has a  \index{propagation of chaos} \emph{propagation of chaos property} as \mbox{$M\to \infty$} for
 a sequence of initial states that are exchangeable, see Proposition \ref{EBAW-chaos1} below.
 If the empirical distributions
 of  $\textbf{Y}^M(0)$ converge to the distribution with weights $\mathbf v_0= (v_0^0,v_0^\eta, v_0^1)$ as $M\to \infty$, then, for each $t>0$,  $\textbf{Y}^M(t)$ converges to the distribution with
 weights $\mathbf v_t = (v_t^0,v_t^\eta, v_t^1)$ given by the solution of the dynamical system \index{ordinary differential equation}
 \begin{align}\label{EBAW-dynsys}
\dot v^0 & = (1-\eta)v^{\eta} - 2 r \eta v^0 (v^1 + \eta v^{\eta}) \nonumber \\
\dot v^{\eta} & = -v^{\eta} + 2 r (\eta^2  v^0 v^{\eta} + (1-\eta)^2 v^1 v^\eta +  v^0 v^1)\\
\dot v^1 & = \eta v^{\eta} - 2 r (1- \eta) v^1 (v^0 +  (1- \eta) v^\eta) \nonumber
 \end{align}
 with initial value $\mathbf v_0$,
 see \cite[Corollary~2.8]{EBAW-PoWa19+}.
 The system leaves the unit simplex $\Delta^3:= \{(z_0,z_\eta,z_1) \in[0,1]^3 : z_0+z_\eta+z_1=1\}$  invariant (as it must).

Here is a brief intuitive explanation of \eqref{EBAW-dynsys}, exemplarily for its first equation. The first term on the r.h.s. of that equation arises from  host replacement, as a sum of what happens when a host of type $0$, type $\eta$ or type $1$ dies. The corresponding contributions to $\dot v^0$ are $-v^0(v^1+\eta v^\eta)$, $v^\eta(v^0+(1-\eta)v^\eta)$ and $v^1(v^0+(1-\eta)v^\eta)$, and these add up to  $(1-\eta)v^\eta$. The second term on the r.h.s. of the first  equation in \eqref{EBAW-dynsys} arises from  reinfection  and corresponds to the rate in the third line of \eqref{EBAW-Yrates}, see also the explanation there.

 The equilibria of \eqref{EBAW-dynsys} and their stability are analysed in
\cite[Proposition~2.11]{EBAW-PoWa19+}. In particular, there is an equilibrium in the interior of $\Delta^3$, which  is  globally stable, if and only if
\begin{align}\label{EBAW-stablecond}
r > \max\left\{ \frac{ 2 \eta-1}{2 \eta(1-\eta)^2}, \frac{1-2 \eta}{2(1-\eta)\eta^2} \right\}.
\end{align}

In the limit $M\rightarrow \infty$, the process of type-$A$ parasite frequencies in a typical host turns out to be a $\{0,\eta, 1\}$-valued Markov process $V=(V(t))_{t\ge 0}$  defined as follows.

\begin{definition}[Evolution of a typical host in the limit $M\to \infty$]\label{EBAW-defV}
For a given $\mathbf v^{}_0 \in  \Delta^3$, let
$\mathbf v = (\mathbf v^{}_t)^{}_{t \geq 0}$ be defined by \eqref{EBAW-dynsys}. We specify the jump  rates of $V$ as follows: 
At time $t$, the process  $V$ jumps  \\
\phantom{AAAAA} from any state to 
 state  
\begin{center}
\begin{tabular}{lll}
 0  & at rate & $v^0_t + (1-\eta) v_t^\eta$, \\
 1  & at rate & $v_t^1 + \eta v_t^\eta$,
\end{tabular}
\end{center} 
\hspace{1.35cm} from state 0  to state  
 $\eta$   at rate  $2 r \eta (v_t^1 +\eta v_t^\eta)$, and\\
\phantom{AAAAA} from state 1 to state $\eta$  at rate
 $2 r (1-\eta)(v_t^0 + (1-\eta)
 v_t^\eta).$ 
 \end{definition}
 The second part of \cite[Corollary 3.4]{EBAW-PoWa19+} gives
 \begin{proposition} \label{EBAW-lawV}
 Let $\mathbf v$ and $V$ be as in Definition \eqref{EBAW-defV}. Then
 \[\mathbb P(V(t) = a) = v_t^a,\quad a\in   \{0,\eta, 1\}.\]
 \end{proposition}
 This confirms that the dynamics of $V$  is of {\em Vlasov-McKean} type: the jump rates of $V$ at time $t$ react on the distribution of $V(t)$. This goes along with the underlying \index{mean field} {\em mean-field} situation. 
 
 Notably, there is a \emph{graphical representation} \index{graphical representation} of the process $(V(t))_{t\ge 0}$ in terms of \index{trees!nested} nested trees $(\mathcal T_t)_{t\ge 0}$, which does not require a prior analytic construction of $\mathbf v$ but rather gives a probabilistic representation of it. This representation is in the spirit of \index{ancestral graph} ancestral graphs in the absence of coalescences, see \cite[Thm.~2]{EBAW-BCH18} or \cite[Prop.~2]{EBAW-BaWa18}.  The leaves of $\mathcal T_t$  are coloured independently according to the distribution with weights~$\mathbf v_0$, with the colours being transported in an interactive way from the leaves to the root; the random state $V(t)$ is then the colour of the root of $\mathcal T_t$.
Let us first describe the construction of $\mathcal T_t$, following  \cite[Definition 3.1]{EBAW-PoWa19+}.

 \begin{figure}[h]
 \includegraphics[height=4.5cm, width=8cm]{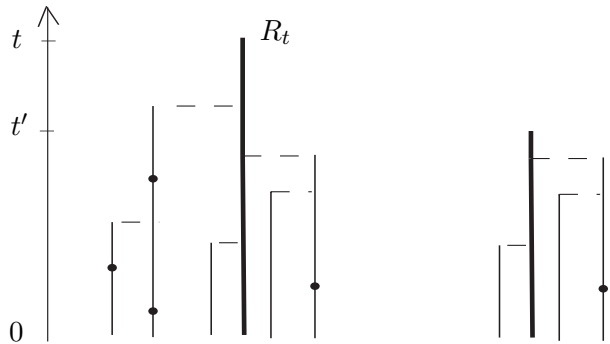}
 \caption {\index{graphical representation} Nested trees in the graphical representation of $V$. Left: A realisation of $\mathcal T_t$. The root $R_t$ is situated at time $t$, the leaves are at time $0$. The distinguished line is drawn bold. HR events along the lines different from the distinguished one are indicated as dots. Branches incoming to the distinguished line at HR events are drawn to the right   of the continuing branch, incoming branches at PER  events (along any line) are drawn to the left of the continuing branch. 
 Right: The corresponding realisation of $\mathcal T_{t'}$ for a $t' < t$, nested into $\mathcal T_t$.}
 \label{EBAW-FigTT}
 \end{figure}
 
\index{graphical representation}
  A~single (distinguished) line starts from the root $R_t$ of $\mathcal T_t$  backwards
in time, this is the downward direction in the (schematic) Figure \ref{EBAW-FigTT}. 
The growth of the tree in downward direction is defined via the splitting rates of its lines. 
Namely, each line is hit by \index{host replacement} {\it host replacement}
({HR}) {\it events} at rate 1 and \index{reinfection} {\it potential effective reinfection} {(PER)} {\it events} at rate~$2r$. At each such event, the line  splits into two branches, the {\rm continuing} and the {\rm incoming} one
(where `incoming'' refers to the direction from the leaves to the root).   Whenever the distinguished line is hit by an HR event, we keep both branches in the tree
and designate the continuing branch as the continuation of the distinguished line. In order to discriminate between PER and HR events along the distinguished line, we draw the incoming line at PER events  to  the left and at at HR events to the right of the distinguished line.
Whenever a line different from the distinguished one is hit by an HR  event, we discard \index{pruning} (or \emph{prune})
the continuing branch and keep only the incoming one; but we keep the event in mind by placing a dot on the continuing line. At a PER event (to \emph{any} line),  we keep both
the incoming and the continuing branches.

Next we describe the transport of the colours that result from an independent colouring of the leaves of $\mathcal T_t$ according to $\mathbf v_0$.   
Assume  an HR  event occurs at time~$\tau$.  If the incoming branch at time $\tau-$  is in state $0$ or~$1$,
then the continuing branch  
takes the state of the incoming branch. If the incoming branch at time $\tau-$ is in state $\eta$, then the state of the continuing
branch at time $\tau$ is decided by a coin toss: it takes the state $1$ or $0$ with probability $\eta$ or $1-\eta$, respectively.  Intuitively, this coin toss decides which type is transmitted (type $A$ with probability $\eta$ and type~$B$ with probability $1-\eta$). 

At a PER event (occurring at time $\tau$, say),  the state of the continuing branch is decided via at most two  independent coin tosses, each with success probability~$\eta$. If, for example, at time $\tau-$  the incoming branch is in state $\eta$ and the continuing branch is in state $0$, then at time $\tau$ the state of the continuing branch changes  to $\eta$ with probability $\eta^2$, and remains in~$0$ with probability $1-\eta^2$. Intuitively,  the first coin toss decides which type is transmitted, and the second coin toss decides whether the \index{reinfection} reinfection is effective. This second coin toss decreases the rate $2r$ of PER events  to the host-state dependent rate of effective reinfection events.  For the rules for the other combinations of states at the incoming and continuing branches, we refer to \cite[Definition 3.1]{EBAW-PoWa19+}.

In this way, given the tree $\mathcal T_t$ and the realisations of the coin tosses indexed by the HR and  PER events along its lines, the states of the leaves are propagated in a deterministic way into the state $C_t$ of the root. As indicated Figure \ref{EBAW-FigTT}, $(\mathcal T_t, C_t)$ can be coupled for various $t$ by nesting the trees in an obvious way.
\begin{proposition}[Graphical construction of the state evolution of a typical host  {\cite[Corollary 3.4]{EBAW-PoWa19+}}] 
\index{graphical representation} For $\mathbf v_0 \in \Delta^3$, 
let $(\mathcal T_t, C_t)_{t\ge 0} $ be constructed as  described above. Then $(C_t)_{t\ge 0}$ has the same distribution as the process $\big (V(t) \big )_{t\ge 0}$ specified in Definition~\ref{EBAW-defV}.
\end{proposition}
This graphical approach is  instrumental for proving the next result als well as  Theorems \ref{EBAW-TfiniteM} and \ref{EBAW-MF}.
\begin{proposition}[Propagation of chaos {\cite[Proposition 2.7]{EBAW-PoWa19+}}] \label{EBAW-chaos1} 
\index{propagation of chaos} 
  Assume  \[\frac 1M \sum_{i=1}^M \delta_{Y_i^M(0)}\to v_0^0 \delta_0 +v_0^\eta \delta_\eta + v_0^1 \delta_1\]
  as $M\to \infty$ for some $\mathbf v_0=(v_0^0, v_0^\eta, v_0^1) \in \Delta^3$.
  Moreover, assume that the initial states $Y^M_1(0), ...,$ $Y^M_M(0)$ are exchangeable, i.e.~arise through
   sampling without replacement from their empirical distribution (given the latter). Then, for each $\overline t > 0$, the random
  paths $Y^M_i = (Y^M_i(t))_{0\le t\le \overline t}$, $i=1,\ldots,M$,  are \index{exchangeable} exchangeable. Furthermore,   for each $k \in \mathbb N$  as $M\to \infty$, one has
 \[(Y^M_1, \ldots, Y^M_k) \to (V_1, \ldots, V_k)\]
  in distribution with respect to  the \index{Skorokhod topology} Skorokhod $\mathrm J_1$-topology, where $V_1, \ldots, V_k$ are i.i.d.~copies  of the process
  $V=(V(t))_{0\le t\le \overline t}$ specified in Definition \ref{EBAW-defV}.
\end{proposition}

\smallskip

\paragraph{{\bf Joint limit:} $\boldmath{M=M_N \rightarrow \infty}$ for $\boldmath{N\rightarrow \infty}$} \label{EBAW-secMdepN}
In analogy to Proposition \ref{EBAW-chaos1}, propagation of chaos can also be shown  in the case of a joint limit of $N$ and $M$ to $\infty$, i.e. 
$M= M_N$ and $M_N \rightarrow \infty$ for $N\rightarrow \infty$. This is the topic of the next theorem. 

\begin{theorem}[Propagation of chaos {\cite[Theorem 2]{EBAW-PoWa19+}}]\label{EBAW-MF}
\index{propagation of chaos}
 Let Assumptions $(\mathcal{A})$ be valid. Assume that, for any $N$, the initial states 
  $ X^{N,M_N}_1(0),...,$ $ X^{N,M_N}_{M_N}(0)$ are \index{exchangeable} exchangeable (i.e. arise via sampling without replacement from their empirical distribution $\mu^N_0$). For $M=M_N\to \infty$ as $N\to \infty$,  assume that $\mu_0^N$ converges weakly as $N\to \infty$ to
 a distribution $\pi$ on $\{0\}\cup[\alpha, 1-\alpha]\cup\{1\}$ for some $\alpha > 0$. 
Let $0 <\underline t  <\overline t< \infty$  and $k \in \mathbb N$. On the time interval $[\underline t ,\overline t]$, the processes $X_1^{N,M_N}, \ldots, X_k^{N,M_N}$ then converge, as \mbox{$N\to \infty$}, to $k$~i.i.d.~copies
  of the process $V$ specified in Definition \ref{EBAW-defV}. Here the distribution of $V(0)$ has  weights $\pi(\{0\}), \pi([\alpha, 1-\alpha]), \pi(\{1\})$, and convergence is in distribution with
  respect to the \index{Skorokhod topology} Skorokhod $\mathrm M_1$-topology.
 \end{theorem}
This allows to derive the following result on the asymptotics of the empirical distributions $\mu^N_t$ of $\big(X^{N,M_N}_i(t), \, i=1,\ldots, M_N\big)$ as $N \to \infty$.
\begin{corollary}[Law of large numbers {\cite[Corollary 2.10]{EBAW-PoWa19+}}]\label{EBAW-EmpDist}
\index{law of large numbers}
 (i) In the situation of Theorem \ref{EBAW-MF},   the sequence of $\mathcal{M}_1(D([\underline t,\overline t]; [0,1])$-valued random variables
 $\mu^N$ converges in distribution (w.r.t. the \index{weak topology} weak topology) to 
the distribution of $V$ as $N\to \infty$, where  the space $D([\underline t,\overline t]; [0,1])$ is equipped with the \index{Skorokhod topology}  Skorokhod $\mathrm M_1$-topology. \\
(ii)   For $t >0$, the sequence of $\mathcal{M}_1([0,1])$-valued random variables
 $\mu^N_t$ converges in distribution (w.r.t. the weak topology) to $v^0_t\delta_0 + v^\eta_t\delta_\eta+ v^1_t\delta_1$ as $N\to \infty$, where $\mathbf v = (v^0, v^\eta, v^1)$ is the solution of \eqref{EBAW-dynsys} with initial condition $\big ( \pi(\{0\}), \pi([\alpha, 1-\alpha]), \pi(\{1\}) \big )$.
 \end{corollary}

\subsection{Maintenance of a polymorphic state} \label{EBAW-poly}
\index{polymorphic}
 For large $N$ and $M$, the weights of the empirical frequencies $\mu_t^N$  are close to the solution of the dynamical system \eqref{EBAW-dynsys} by Corollary \ref{EBAW-EmpDist}.  Hence --- once a state close to the stable equilibrium state of \eqref{EBAW-dynsys} is reached --- both parasite types $A$ and $B$
are maintained in the population for a long time. Indeed, an asymptotic lower bound for this time, which is `almost exponential' in $M_N$, is given in \cite[Theorem 3ii)]{EBAW-PoWa19+}. However, with $N$ being finite, eventually one of the types goes to  \index{fixation} fixation and the population enters 
a  \index{monomorphic} monomorphic state with all hosts  infected with either   type $A$ or   $B$ only.  

To formally overcome this problem of ultimate fixation, we now enrich our model by allowing, in addition to the rates specified in  Definition \ref{EBAW-jumpratesMN}, a \index{mutation!two-way} two-way mutation for the parasites  at rate $u^{}_N$ per parasite generation. Then 
\begin{equation}\label{EBAW-popmutrate}
\theta_N:= u^{}_N N M_N g^{}_N
\end{equation}
is the {\em population mutation rate}, i.e.~the total rate at which parasites mutate in the total host population on the host time
scale. 
If $\theta_N = o(r^{}_N)$,  the  rate at which a type is transmitted by \index{reinfection} reinfections
is much larger than the mutation rate to this type, even if this type is
retained only in a single host (at around 
the equilibrium frequency). The dynamical system arising as the limiting evolution of $X^{N,M}$ as $N\to \infty$ is then not perturbed by mutations. 

Even though most mutations away from a monomorphic population will get lost due to fluctuations, the assumed recurrence of the mutations will eventually turn  a  monomorphic host population into a polymorphic one. The condition  
\begin{equation}\label{EBAW-brachcoup}
r>\max\left\{ \frac{\eta}{2 (1-\eta)^2}, \frac{1-\eta}{2 \eta^2}\right\},
\end{equation}
which is stronger than \eqref{EBAW-stablecond}, allows for a \index{coupling} coupling with a supercritical \index{branching process} branching process that estimates from below the number of hosts infected with the  currently-rare parasite type. Under this condition together with Assumptions $(\mathcal A)$,  \cite[Theorem 3i)]{EBAW-PoWa19+}  gives an asymptotic 
upper bound (in terms of $\theta_N$, $s^{}_N$ and $M^{}_N$) on the time at which, with high probability (i.e.~with a probability that tends to 1 as $N\rightarrow \infty$), the empirical distribution of the host's states reaches a small neighborhood of the stable fixed point of the dynamical system \eqref{EBAW-dynsys}, when the particle system is   started from a monomorphic state.

A comparison of the two bounds in parts (i) and (ii) of  \cite[Theorem 3]{EBAW-PoWa19+}  shows that, as long as \mbox{$\theta_N = o(r^{}_N)$} and $\theta_N$ obeys a mild asymptotic lower bound,  
the proportion of time  the population spends in a \index{monomorphic} monomorphic state is negligible relative to the time it spends in a \index{polymorphic} polymorphic state. In fact the required lower bound on $\theta_N$ turns out to be subexponentially small in the host population size. From a modelling perspective, it seems important that the lower bound is this small. Indeed, the genetic variability we are modelling 
is found in coding regions. Types $A$ and  $B$ represent different genotypes/alleles of the same gene (e.g.\ in HCMV there exist two genotypes for the 
region UL 75; they are separated by one deletion (removing an amino acid) and  8 amino acid substitutions, requiring at least 8 non-synonymous point mutations). Since no
`intermediate genotypes' have been found,
it is likely that a fitness valley lies between the two genotypes, see \cite{EBAW-PoGo19+} for more details
on the biological motivation. 

\subsection*{Acknowledgement}
We thank Cornelia Pokalyuk for comments that helped to improve the presentation, Sebastian Probst for help with figures, and an anonymous referee for thoroughly reading the manuscript and for insightful questions and suggestions.


\begin{thebibliography}{99}\itemsep=2pt

\bibitem{EBAW-Ba08}
E.~Baake,
The Luria-Delbr\"{u}ck experiment: are mutations spontaneous or directed?,
\textit{EMS Newsletter} \textbf{69} (2008), 17--20.

\bibitem{EBAW-BGPW19}
E.~Baake, A.~Gonz\'{a}lez Casanova, S.~Probst, and A.~Wakolbinger,
Modelling and simulating Lenski's long-term evolution experiment,
\textit{Theor. Popul. Biol.} \textbf{127} (2019), 58--74.

\bibitem{EBAW-BaWa18}
E.~Baake and A.~Wakolbinger,
Lines of descent under selection,
\textit{J. Stat. Phys.} \textbf{172} (2018), 156--174.

\bibitem{EBAW-BCH18}
E.~Baake, F.~Cordero, and S.~Hummel,
A probabilistic view on the deterministic mutation-selection equation: dynamics, equilibria, and ancestry via individual lines of descent,
\textit{J. Math. Biol.} \textbf{77} (2018), 795--820.

\bibitem{EBAW-RBPP20}
R.~Backofen and P.~Pfaffelhuber,
The population genetics of the CRISPR-Cas system in bacteria,
\emph{in preparation}.

\bibitem{EBAW-BGPoW19}
F. Boenkost, A.~Gonz\'{a}lez Casanova, C. Pokalyuk, and A. Wakolbinger, \\Haldane's formula in Cannings models: The case of moderately weak selection, \emph{preprint}, \texttt{arXiv:1907.10049}

\bibitem{EBAW-Ca10}
M.J. Cannon, D.S. Schmid and T.B. Hyde, Review of cytomegalovirus seroprevalence and demographic characteristics associated with infection, 
\textit{Reviews in Medical Virology}
 \textbf{20} (2010), 202--213.

 
\bibitem{EBAW-Cha06}
N.~Champagnat,  A microscopic interpretation for adaptive dynamics trait substitution sequence models,
\textit{Stochastic Process. Appl.}
 \textbf{116} (2018), 1127--1160.

\bibitem{EBAW-Chevin11}
L.M.~Chevin,
On measuring selection in experimental evolution,
\textit{Biology Letters} \textbf{7} (2011), 210--213.

\bibitem{EBAW-DeFi07}
M.M.~Desai and D.~S.~Fisher,
Beneficial mutation-selection balance and the effect of linkage on positive selection,
\textit{Genetics} \textbf{176} (2007), 1759--1798.


\bibitem{EBAW-Da18}
D.A.~Dawson,  Multilevel mutation-selection systems and set-valued duals,
\textit{J.Math.Biol.}
 \textbf{76} (2018), 295--378.


\bibitem{EBAW-Du08}
R.~Durrett,
Probability Models for DNA Sequence Evolution,
2nd ed., Springer, New York, 2008.

\bibitem{EBAW-Ewens04}
W.J.~Ewens,
Mathematical Population Genetics,
2nd ed., Springer, New York, 2004.

\bibitem{EBAW-EWK07}
A.~Eyre-Walker and P.D.~Keightley,
The distribution of fitness effects of new mutations,
\textit{Nature Reviews Genetics} \textbf{8} (2007), 610--618.

\bibitem{EBAW-PuGoe11}
E. Puchhammer-St{\"o}ckl  and I. G{\"o}rzer, 
Human cytomegalovirus: an enormous variety of strains and their possible clinical significance in the human host,
\textit{Future Medicine} \textbf{6} (2011), 259--271.

\bibitem{EBAW-Fi1918}
R.~Fisher, 
The correlation between relatives on the supposition of Mendelian inheritance, 
\textit{Phil.~Trans.  R. Soc.  Edinburgh} \textbf{52} (1918), 399--433.

\bibitem{EBAW-GL98}
P.J. Gerrish and R.E. Lenski,
\newblock The fate of competing beneficial mutations in an asexual population,
\newblock Genetica \textbf{102/103} (1998), 127--144.


\bibitem{EBAW-Gill84}
J.H. Gillespie,  Molecular evolution over the mutational
landscape, 
\textit{Evolution} \textbf{38} (1984), 1116--1129.


\bibitem{EBAW-GKWY16}
A.~Gonz\'{a}lez Casanova, N.~Kurt, A.~Wakolbinger,  and L.~Yuan, 
An individual-based model for the Lenski experiment, and the deceleration of the relative
fitness,
\textit{Stoch. Process. Appl.} \textbf{126} (2016), 2211--2252.

\bibitem{EBAW-KTP09}
S.~Kryazhimskiy, G.~Tka\u{c}ik, and J.~B.~Plotkin, 
The dynamics of adaptation on correlated fitness landscapes,
\textit{Proc.~Natl.~Acad.~Sci.~USA} \textbf{106} (2009), 18638--18643.

\bibitem{EBAW-Le19} R. E. Lenski,  The E. coli long-term experimental evolution project site.\\ http://myxo.css.msu.edu/ecoli (2019)

\bibitem{Lenski91}
R.~Lenski, M.R.~Rose, S.~Simpson, and S.C.~Tadler,
Long term experimental evolution in \emph{Escherichia coli} I. Adaptation and divergence during 2000 generations,
\textit{Amer. Nat.} \textbf{138} (1991), 1315--1341.

\bibitem{EBAW-LT94}
R. E. Lenski and M. Travisano, 
Dynamics of adaptation and diversification: a 10,000-generation experiment with bacterial populations,
\textit{Proc. Natl. Acad. Sci. U.S.A.} \textbf{91} (1994), 6808--6814.


\bibitem{EBAW-LuDe43}
S. Luria and M. Delbr\"uck, 
Mutations of bacteria from virus sensitivity to virus resistance, 
\textit{Genetics} \textbf{28} (1943), 491--511.

\bibitem{EBAW-LuMa17}
S. Luo and C. Mattingly,
Scaling limits of a model for selection at two scales,
\textit{Nonlinearity} \textbf{30} (2017).

\bibitem{EBAW-McCS14}
D. M. McCandlish and A. Stoltzfus,
Modelling evolution using the probability of fixation: History and implications,
\textit{The Quarterly Review of Biology} \textbf{89} (2014), 225--252.

\bibitem{EBAW-MS76}
J. Maynard Smith,
What determines the rate of evolution?
\textit{Amer. Nat.} \textbf{110} (1976), 331--338.

\bibitem{EBAW-MoLaetal17}
  D.H.~Morris,  K.~Gostic, S.~Pompei, T.~Bedford, M.~{\L}uksza, R.A.~Neher,
  B.T.~Grenfell, M.~L\"assig, J.W.~McCauley,
  Predictive modeling of influenza shows the promise of applied evolutionary
  biology,
  \textit{Trends in Microbiology} \textbf{26} (2018), 102--118.



\bibitem{EBAW-Orr03}
H.A. Orr, The distribution of fitness effects among
beneficial mutations, 
\textit{Genetics} \textbf{163} (2003), 1519--1526.


\bibitem{EBAW-PW08}
Z. Patwa and L.M. Wahl, 
The fixation probability of beneficial mutations, 
\textit{J. R. Soc. Interface}  \textbf{5} (2008), 1279--1289.

\bibitem{EBAW-PoGo19+}
C.~Pokalyuk and I.~G\"orzer,
Diversity patterns in parasite populations capable for persistence and reinfection with a view towards the \emph{human cytomegalovirus},
preprint, \texttt{bioRxiv doi: 10.1101/512970}


\bibitem{EBAW-PoPf13}
C. Pokalyuk and P. Pfaffelhuber,
The ancestral selection graph under strong directional selection, 
\textit{Theor. Popul. Biology}  \textbf{87} (2013), 25--33.


\bibitem{EBAW-PoWa19+}
C.~Pokalyuk and A.~Wakolbinger,
Maintenance of diversity in a parasite population capable of persistence and reinfection, 
\textit{Stoch. Processes Appl.}, in press;
\texttt{arXiv:1802.02429}


\bibitem{EBAW-Sanjuan10}
R. Sanju\'an,
Mutational fitness effects in RNA and single-stranded DNA viruses: common patterns revealed by site-directed mutagenesis studies,
\textit{Phil. Trans. R. Soc. B} \textbf{365} (2010), 1975--1982.

\bibitem{EBAW-Skorokhod56}
A. V. Skorokhod,
Limit Theorems for Stochastic Processes,
\textit{Theory Probab. Appl.} \textbf{1} (1956), 261--290.


\bibitem{EBAW-WSAT20}
W.~Stephan and A.~Tellier,
Stochastic processes and host-parasite coevolution: linking coevolutionary dynamics and DNA polymorphism data, \emph{in preparation}.

\bibitem{EBAW-WRL13}
M.J.~Wiser, N.~Ribeck, and R.E. Lenski,
Long-term dynamics of adaptation in asexual populations,
\textit{Science} \textbf{342} (2013), 1364--1367.






\end{thebibliography}
\end{document}